\documentclass[a4paper,11pt]{article}

\usepackage{pos}
\usepackage{booktabs}
\usepackage{tabularx}
\usepackage{dsfont}
\usepackage{braket}
\usepackage{amsmath}
\usepackage{graphicx}
\usepackage[section]{placeins}
\newcommand{\qhat}{\ensuremath{\hat{q}} }
\newcommand{\phat}{\ensuremath{\hat{p}} }

\newcommand{\bhat}{\ensuremath{\hat{b}} }
\newcommand{\bhatdag}{\ensuremath{\hat{b}^\dag} }

\newcommand{\La}{\ensuremath{\Lambda} }

\newcommand{\fig}[1]{Fig.~\ref{#1}}
\newcommand{\tab}[1]{Table~\ref{#1}}

\title{Simulating Supersymmetric Quantum Mechanics Using Variational Quantum Algorithms}

\author*{John Kerfoot}
\author{David Schaich}
\author{Emanuele Mendicelli}

\affiliation{Department of Mathematical Sciences, University of Liverpool, Liverpool L69 7ZL, United Kingdom}

\emailAdd{john.kerfoot@liverpool.ac.uk}
\emailAdd{david.schaich@liverpool.ac.uk}
\emailAdd{e.mendicelli@liverpool.ac.uk}

\abstract{The study of spontaneous supersymmetry breaking (SSB) on the lattice is obstructed by a severe sign problem. Quantum computing provides a promising alternative approach. In particular, properties of supersymmetry relate SSB to the ground-state energy, which can be probed using hybrid quantum--classical algorithms such as the variational quantum eigensolver (VQE). In this work we present VQE analyses for supersymmetric quantum mechanics with various superpotentials. A key new feature is an adaptive ansatz construction algorithm that reduces the number of variational parameters within our ans{\"a}tze. This lowers the resource burden on both the classical optimizer and the noisy quantum processor, thereby improving the feasibility of these calculations in the NISQ era. Additionally, we present preliminary VQE results obtained from real IBM quantum devices, highlighting accuracy, resource constraints, and computational cost, both with and without the application of error mitigation techniques.}

\FullConference{The 42nd International Symposium on Lattice Field Theory (LATTICE2025)\\
2-8 November 2025\\
Tata Institute of Fundamental Research, Mumbai, India\\}

\begin{document}
\maketitle

\section{Introduction}
Supersymmetry extends the Poincar\'e spacetime symmetry group by adding fermionic generators that relate bosonic and fermionic fields. This increased symmetry offers a vast array of applications, ranging from hypothetical extensions of the Standard Model, to deepening our understanding of the foundational structures of quantum field theories. Lattice methods have become a powerful non-perturbative method for studying a wide variety of strongly coupled supersymmetric theories --- see Ref.~\cite{Schaich:2022xgy} for a recent review.  In this work, we investigate spontaneous supersymmetry breaking (SSB) in 0+1D supersymmetric quantum mechanics (SQM).

A standard approach for studying lattice field theories is through Monte Carlo importance sampling of the Euclidean path integral. However, for real-time dynamics the weight becomes complex, making importance sampling no longer applicable. Additionally, SSB implies a vanishing Witten index, which corresponds to a vanishing Euclidean partition function. This implies a severe sign problem and poses a significant challenge for standard Monte Carlo methods. This motivates studying SSB within the Hamiltonian formalism where sign problems are avoided.

Quantum computers provide a natural method for studying these systems. On a classical device the exponentially growing Hilbert space would make studying SSB in the Hamiltonian formalism infeasible however quantum computing makes accessing this Hilbert space potentially viable using polynomial resources.
In the current noisy intermediate-scale quantum (NISQ) computing era, hardware noise and circuit depth are key limiting factors in quantum applications. Due to this, NISQ applications must be designed to be robust to noise. Variational methods have been an important focus of recent research~\cite{Farrell:2024fit}, with applications spanning a variety of fields from quantum chemistry and biology to particle and high energy physics. 

Because the study of SSB is highly linked to the energy spectrum of the system, and in particular the groundstate energy, we explore the use of the variational quantum eigensolver (VQE)~\cite{Peruzzo:2013bzg}. Since this algorithm is highly dependent on the choice of parametrized quantum circuit (ansatz) we demonstrate how an adaptive approach can be used to construct hardware-efficient and problem-tailored ans\"atze. In Section~\ref{SQM} we introduce the SQM Hamiltonian and the three different superpotentials we will consider in this work. In Section~\ref{RD} we discuss how we regulate the infinite-dimensional bosonic Hilbert space as well as how we digitize and map the Hamiltonian to a qubit device. In Section~\ref{AVQE} we introduce the Adaptive-VQE algorithm and demonstrate how truncating ans\"atze can be beneficial when considering noisy hardware.
We recently reported these results in Ref.~\cite{kerfoot2025}.
In Section~\ref{HR} we go beyond Ref.~\cite{kerfoot2025} and present a brief summary of preliminary results from running circuits on real IBM devices before concluding and discussing future directions in Section~\ref{Conclusion}.

\section{Supersymmetric Quantum Mechanics}\label{SQM}

The SQM theory involves a single fermionic and bosonic degree of freedom, both fixed at a single spatial site while being able to evolve in continuous time.
The Hamiltonian can be written as
\begin{equation}\label{eq:H_SQM}
H = \frac{1}{2}\left( \phat^2 + [W'(\qhat)]^2 - W''(\qhat) \left[\bhatdag, \bhat\right]\right).
\end{equation}
The momentum and position coordinate operators of the boson are represented by the operators \phat and \qhat respectively, which satisfy the standard commutation relation $[\qhat,\phat]=i$. The fermionic annihilation and creation operators are represented by the operators \bhat and \bhatdag and satisfy the standard canonical anti-commutation relation $\lbrace \bhat, \bhatdag \rbrace = \mathds{1}$. $W(\qhat)$ represents a choice of superpotential with $W'(\qhat)$ representing differentiation with respect to the position operator $\qhat$. Looking at the final term in Eq.~\ref{eq:H_SQM} one can see the interactions between the bosonic and fermionic degrees of freedom are governed by the choice of superpotential. In this work we consider the three superpotentials shown in Table~\ref{tab:superpotential}, setting $m=g=\mu=1$ for simplicity. 
Since our choice of superpotential determines the interactions, it also governs whether supersymmetry is preserved or spontaneously broken~\cite{Cooper:1994eh, Woit:2017vqo}.
Of the three superpotentials we consider, we expect the Harmonic Oscillator (HO) and Anharmonic Oscillator (AHO) to preserve supersymmetry and the Double Well (DW) to lead to SSB.

We can represent the Hamiltonian $H= \frac{1}{2}\lbrace Q, Q^{\dag} \rbrace$ in terms of supercharges
\begin{align}\label{eq:Qs}
  Q & = \bhat \left( i \phat + W'(\qhat) \right) &
  Q^{\dag} & = \bhatdag \left( -i \phat + W'(\qhat) \right) &
  Q^2 & = (Q^{\dag})^2 = 0.
\end{align}
A consequence of these conditions is that all eigenstates will have non-negative energy. Additionally, a zero-energy groundstate is only possible if the state preserves supersymmetry and therefore is annihilated by both supercharges, i.e.\ $Q^{\dag} \ket{\Phi} = Q \ket{\Phi} = 0$. Conversely, if we see a pair of positive-energy groundstates then this would indicate supersymmetry is spontaneously broken. Therefore, the preservation or spontaneous breaking of supersymmetry for a given superpotential
can be determined by measuring the ground-state energy of the system. 

As an aside, while the relation between spontaneous supersymmetry breaking and the ground-state energy is generic, the special case of SQM with a polynomial superpotential is simple enough that the preservation or spontaneous breaking of supersymmetry can be read off directly from the superpotential~\cite{Woit:2017vqo, Gendenshtein:1985hgo}. Specifically, if the polynomial degree of the superpotential is even, then a normalizable solution to $Q^{\dag} \ket{\Phi} = Q \ket{\Phi} = 0$ can be found, implying preserved supersymmetry.

\begin{table}[!htbp]
\centering
\small
\setlength{\tabcolsep}{8pt}

\begin{tabularx}{\linewidth}{>{\centering\arraybackslash}X
                            >{\centering\arraybackslash}X
                            >{\centering\arraybackslash}X}
\toprule
Harmonic Oscillator (HO) & Anharmonic Oscillator (AHO) & Double Well (DW) \\
\addlinespace
$W(\qhat)=\frac{1}{2}m\qhat^2$ &
$W(\qhat)=\frac{1}{2}m\qhat^2 + \frac{1}{4} g \qhat^4$ &
$W(\qhat)=\frac{1}{2}m\qhat^2 + g\!\left(\frac{1}{3}\qhat^3 + \mu^2 \qhat\right)$ \\
\addlinespace
Supersymmetric & Supersymmetric & Spontaneously broken \\
\bottomrule
\end{tabularx}

\caption{The three superpotentials used in this work, labelled by whether they are expected to preserve supersymmetry or allow SSB.}
\label{tab:superpotential}

\end{table}

\section{Regularization \& Digitization}\label{RD}
In order to encode our SQM Hamiltonian to a qubit device we must first regularize the infinite-dimensional bosonic Hilbert space and represent it with finite resources. We do this by introducing a cutoff \La on the number of allowed bosonic modes.
We encode the boson using the Fock basis meaning each computational basis state is the occupation number and our position and momentum operators are
\begin{equation}\label{eq:pq}
\begin{aligned}
\qhat &\doteq \frac{1}{\sqrt{2m}}\;\scalebox{0.5}{$\begin{bmatrix}
0 & \sqrt{1} & 0 & \cdots & 0\\
\sqrt{1} & 0 & \sqrt{2} & \cdots & 0\\
0 & \sqrt{2} & \ddots & \ddots & 0\\
\vdots & \vdots & \ddots & 0 & \sqrt{\Lambda-1}\\
0 & 0 & 0 & \sqrt{\Lambda-1} & 0
\end{bmatrix}$}
\qquad\qquad\quad
\phat &\doteq i\sqrt{\frac{m}{2}}\;\scalebox{0.5}{$\begin{bmatrix}
0 & -\sqrt{1} & 0 & \cdots & 0\\
\sqrt{1} & 0 & -\sqrt{2} & \cdots & 0\\
0 & \sqrt{2} & \ddots & \ddots & 0\\
\vdots & \vdots & \ddots & 0 & -\sqrt{\Lambda-1}\\
0 & 0 & 0 & \sqrt{\Lambda-1} & 0
\end{bmatrix}$}.
\end{aligned}
\end{equation}
Since we are using devices based on two-state qubits, we can represent the number of bosonic modes \La in binary using $\Lambda = 2^{N_b}$, where $N_b$ is the number of qubits allocated to the boson. Although regularization is essential for rendering the theory computationally tractable, it introduces unphysical artifacts. In particular, the effects at small \La can explicitly cause supersymmetry breaking in a system where we would not expect it. These artifacts are reduced by removing the truncation $\La \to \infty$. This concretely means that systems with a range of \La must be studied to conclusively determine whether a system spontaneously breaks or preserves supersymmetry.

Mapping the fermion to the qubit device is straightforward since we can use the well-known Jordan--Wigner transformation. In this representation, the absence of a fermion is denoted by the state $\vert 0  \rangle $, and its presence is represented by $\vert 1 \rangle $, while the operators are
\begin{align}
  \label{eq:Jordan_Wigner}
  \bhat & = \frac{1}{2} (X +iY) &
  \bhatdag & = \frac{1}{2} (X - iY) &
  \Rightarrow \; \left[\bhatdag, \bhat\right] & = -Z,
\end{align}
where $X$, $Y$ and $Z$ represent the Pauli gates.

\section{Adaptive-VQE}\label{AVQE}
Studying the energy spectrum of quantum systems is a common application of quantum computing, and variational algorithms have become a favourable approach for solving this task. The VQE~\cite{Peruzzo:2013bzg} is a well-known variational algorithm that iteratively minimizes an objective (or `cost') function. When this objective function is chosen to be the energy, the VQE aims to minimize the energy to the ground state of the system. The VQE has already been demonstrated to be successful for a wide variety of systems; for a comprehensive review, see Ref.~\cite{Tilly:2021jem} and references therein.

A major drawback for the VQE is that the performance of the algorithm is highly dependent on the choice of classical optimizer and parametrized quantum circuit (ansatz). If the ansatz is not expressive enough then it may be impossible to accurately represent the groundstate. On the other hand, increasing expressivity requires deeper circuits, more entangling gates and often causes the circuit to scale poorly with the number of qubits. This can be observed in many hardware-efficient ans\"atze~\cite{HardwareEfficent:2019} and a consequence of increasing expressivity is that the classical optimizer may struggle to resolve the increasing number of variational parameters. Additionally, by increasing circuit depth and gate counts we also introduce additional gate error within a noisy quantum device. 
This issue has sparked recent research into systematic methods of creating problem-tailored ans\"atze using more adaptive approaches. 

The Adaptive Derivative-Assembled Problem Tailored (ADAPT) VQE, and similar methods based on it, have been successfully applied to similar problems~\cite{Grimsley:2018wnd, Grimsley:2022azc, Farrell:2023fgd, Farrell:2024fit, Farrell:2025nkx, Gustafson:2024bww, Shirali:2025cox}. These methods aim to construct an ansatz iteratively by appending, at each step, an operator selected from a pre-defined pool. The choice of operator is determined by how it affects the energy, which for example can be taken as largest-magnitude gradient of the expectation value, the infidelity, or the commutator of the Hamiltonian with the operator. This ensures operators that have the largest effect on the energy are included first and therefore reduces the likelihood of including unnecessary gates.

Motivated and inspired by these methods, we propose a related approach, which we will simply refer to as an Adaptive-VQE (AVQE) algorithm. AVQE retains the core ideas as the ADAPT-VQE algorithm but differs in how we construct the operator pool. Instead of using parametrized time-evolution operators we instead define our operator pool using single-qubit rotations as well as two-qubit controlled rotations.

Operator selection is determined by calculating the largest-magnitude gradient for the expectation value of the Hamiltonian.  Specifically, the AVQE algorithm is performed as follows:
\begin{enumerate}
    \item Define an operator pool \{$\hat{O}$\} as discussed further below.
    \item Initialize the ansatz state $\ket{\psi_{\text{ansatz}}}$ in a chosen basis state.
    \item Measure the gradient of the expectation value of the Hamiltonian for each operator in the pool:
        \begin{equation}\label{eq:ADAPT-VQE}
            \nabla\bra{\psi_{\text{ansatz}}}\hat{O}_{i}(\theta)^\dag\hat{H}\hat{O}_{i}(\theta)\ket{\psi_{\text{ansatz}}}
        \end{equation}
    \item Append to the ansatz the operator that returns the largest gradient.
    \item Perform a VQE using the updated ansatz. If an operator $\hat O_i$ has been optimized in a previous VQE step then use the previous optimal parameter as the initial rotation angle $\theta_i$. For the new operator $\hat O_n$ initialize $\theta_n=0$.
    \item If the change in energy is below a pre-determined threshold then terminate the algorithm. Else return to step 3 with the updated ansatz.
\end{enumerate}

Implementing this algorithm on noisy hardware may be challenging as it is highly sensitive to gradient evaluations which can incur significant errors on a real quantum device. To address this, we instead perform the algorithm classically using statevector simulation on small, tractable systems, with the aim of identifying patterns in the ans\"atze that can then be extrapolated to larger system sizes.

Choosing an operator pool requires a balance between expressivity and resource cost. An operator pool that is too large would increase the number of times Eq.~\ref{eq:ADAPT-VQE} is calculated per step while a pool that is too small may not contain the essential gates required to construct the groundstate. In this work we find the following pool to be effective:
\begin{equation}
    \{{\hat{O}}\} = \{RY, RZ, CRY\}.
\end{equation}

From direct inspection of energy eigenvectors for small $\La$, we can see the ground state for the HO and AHO superpotentials has the fermion qubit in the $\ket{1}$ state. Similarly, for the DW superpotential with $\La \geq 8$ the ground state has the fermion qubit in the $\ket{0}$ state. This is used to motivate our choice of basis state. Our initial states are $\ket{10\dots0}$ for the HO and AHO superpotentials, and $\ket{00\dots0}$ for the DW superpotential with $\La \neq 4$. Utilizing these basis states initializes our statevector with a non-zero overlap with the groundstate eigenvector therefore placing the optimizer in a preferable search space, making it significantly easier for the algorithm to find the correct ground state.

\begin{table}
\centering
\resizebox{\textwidth}{!}{%
\begin{tabular}{lrlclll}
\toprule
$W(\qhat)$ & \La & Basis State & $N_\text{gates}$ & Ansatz & $E_{\text{VQE}}$ & $E_{\text{Exact}}$ \\
\midrule
AHO & 2 & $\ket{10}$ & 1 & RY[$q_1$] & -0.437500 & -0.437500 \\
AHO & 4 & $\ket{100}$ & 1 & RY[$q_1$] & -0.164785 & -0.164785 \\
AHO & 8 & $\ket{1000}$ & 3 & RY[$q_2$], RY[$q_1$], CRY[$q_1$, $q_2$] & 0.032010 & 0.032010 \\
AHO & 16 & $\ket{10000}$ & 7 & RY[$q_2$], RY[$q_3$], RY[$q_1$], CRY[$q_1$, $q_2$], CRY[$q_1$, $q_3$], CRY[$q_2$, $q_3$], CRY[$q_3$, $q_2$] & -0.001167 & -0.001167 \\
AHO & 32 & $\ket{100000}$ & 24 & RY[$q_2$], RY[$q_3$], RY[$q_1$], CRY[$q_1$, $q_2$], CRY[$q_1$, $q_3$], RY[$q_4$], CRY[$q_2$, $q_3$] $\dots$ , RY[$q_3$] & 0.000006 & 0.000006 \\
AHO & 64 & $\ket{1000000}$ & 28 & RY[$q_2$], RY[$q_3$], RY[$q_1$], CRY[$q_1$, $q_2$], CRY[$q_1$, $q_3$], RY[$q_4$], CRY[$q_2$, $q_3$] $\dots$  , RY[$q_4$] & 0.000019 & -0.000000 \\
$\vdots$&&&&&&\\
AHO & \La & $\ket{10\dots0}$ & 4 & RY[$q_2$], RY[$q_3$], RY[$q_1$], CRY[$q_1$, $q_2$] &&\\
\midrule
DW & 2 & $\ket{00}$ & 1 & RY[$q_0$] & 0.357233 & 0.357233 \\
DW & 4 & $\ket{100}$ & 3 & RY[$q_1$], RY[$q_0$], CRY[$q_1$, $q_0$] & 0.906560 & 0.906560 \\
DW & 8 & $\ket{0000}$ & 7 & RY[$q_0$], CRY[$q_0$, $q_1$], RY[$q_2$], RY[$q_1$], CRY[$q_0$, $q_2$], CRY[$q_1$, $q_2$], RY[$q_1$] & 0.884580 & 0.884580 \\
DW & 16 & $\ket{00000}$ & 18 & RY[$q_0$], CRY[$q_0$, $q_1$], RY[$q_2$], RY[$q_1$], RY[$q_3$], CRY[$q_0$, $q_2$], CRY[$q_0$, $q_3$] $\dots$ , RY[$q_2$] & 0.891599 & 0.891599 \\
DW & 32 & $\ket{000000}$ & 23 & RY[$q_0$], CRY[$q_0$, $q_1$], RY[$q_2$], RY[$q_1$], RY[$q_3$], CRY[$q_0$, $q_2$], CRY[$q_0$, $q_3$] $\dots$ , RY[$q_3$] & 0.891634 & 0.891632 \\
DW & 64 & $\ket{0000000}$ & 27 & RY[$q_0$], CRY[$q_0$, $q_1$], RY[$q_2$], RY[$q_1$], RY[$q_3$], CRY[$q_0$, $q_2$], CRY[$q_0$, $q_3$] $\dots$ , RY[$q_5$] & 0.891635 & 0.891632 \\
$\vdots$&&&&&&\\
DW & \La & $\ket{00\dots0}$ & 4 & RY[$q_0$], CRY[$q_0$, $q_1$], RY[$q_2$], RY[$q_1$] &&\\
\bottomrule
\end{tabular}
}
 \caption{Summary of 100 AVQE results for the AHO and DW superpotentials and increasing number of bosonic modes $\La$, using statevector simulation and the COBYQA optimizer. The rotation gates in the `Ansatz' column act on the qubit(s) specified by the number(s) in the brackets. In the case of CRY the control and target qubits are [control,target]. The final two columns compare the best ansatz VQE energy from the 100 runs and the true ground-state energy from exact diagonalization. The final row for each superpotential shows the generalized ansatz extrapolated to arbitrary $\La$.}
 \label{tab:AVQE-Summary}
\end{table}

\begin{figure}
    \centering
    \includegraphics[width=1.0\textwidth]{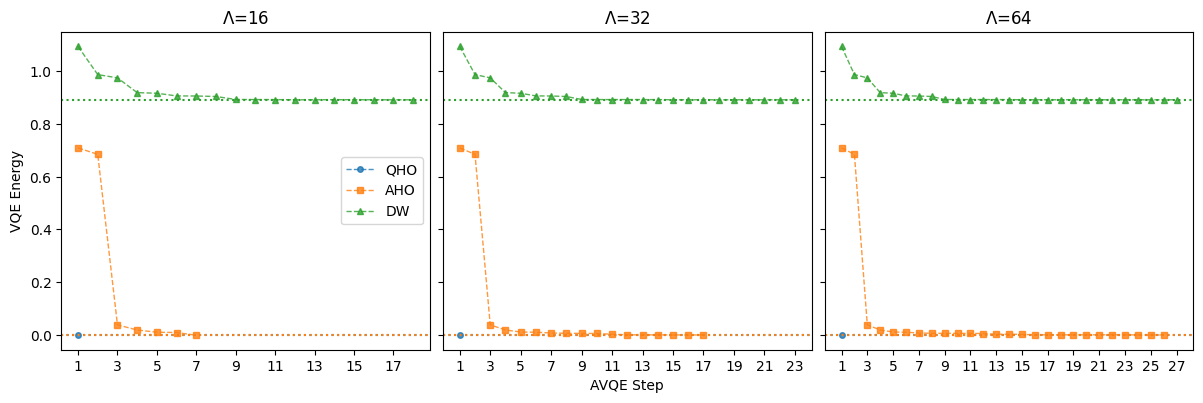}
    \caption{AVQE energies at each step of the algorithm for each superpotential and increasing values of $\La$, using PennyLane's default.qubit statevector simulation and the COBYQA optimizer. Dashed lines indicate the minimum eigenvalue from exact diagonalization.}
    \label{fig:avqe-steps}
\end{figure}

\tab{tab:AVQE-Summary} summarizes AVQE ansatz construction results for the AHO and DW superpotentials with \La ranging from 2 to 64. We omit the HO superpotential for which the results are trivial~\cite{kerfoot2025}. In each case we ran the algorithm 100 times using different random initialization of the classical COBYQA optimizer. The ans\"atze shown under the `Ansatz' column are those that returned the VQE energy closest to the exact value calculated from exact diagonalization of the Hamiltonian. This exact groundstate value is indicated by the $E_{\text{Exact}}$ column and the VQE energy by $E_{\text{VQE}}$. It is possible that better ans\"atze may exist and in fact this appears to become increasingly likely as we consider larger \La where the optimization becomes more challenging. This can be seen in \fig{fig:avqe-steps}: at larger \La the VQE energy decreases sharply in the initial steps but then it appears to plateau and only small improvements occur as more gates are added. This may indicate the optimizer is struggling to converge meaning the best ans\"atze may be smaller than what the algorithm suggests. \tab{tab:AVQE-Summary} is consistent with this idea and shows as we get to larger \La the precision of the VQE energy begins to decline. Again, this is likely a result of the classical optimizer struggling to resolve the growing number of variational parameters together with an increasingly complex Hamiltonian.

Despite this issue we can still observe the systematic construction of the ans\"atze. For the trivial HO superpotential, only a single RY gate on the fermion ($q_n$) is required. This is no surprise since this model is non-interacting and we are representing our boson in the Fock basis. This is the case for all \La and can be easily extrapolated to arbitrary $\La$. There is also a clear pattern in the construction of ans\"atze for the AHO and DW superpotentials. Although the total number of gates in the ansatz, $N_{\text{gates}}$, increases with $\La$, we can see the same initial gates are added for all $\La \ge 16$. Since, by design, at each step the algorithm adds the gates that have the most impact on the energy, we can infer that the initial gates seen in \tab{tab:AVQE-Summary} have the most impact and subsequent gates have diminishing effects. As discussed above, this can also be seen in \fig{fig:avqe-steps}.

When targeting NISQ devices resource cost and quantum noise must be considered. If we choose to utilize a higher number of gates to make the ansatz more expressive then we must also pay the price of having additional gate noise as a result. If we were to use the full circuits returned from the AVQE algorithm then the gate errors from the large circuit sizes may outweigh the energy gain possible from having the more expressive ansatz. We have already discussed how the initial gates have the most impact on the energy and therefore we argue that truncating the ans\"atze can be beneficial when running these circuits on NISQ devices. In Ref.~\cite{kerfoot2025} we compare classical simulation results for the truncated and full ans\"atze and we find truncating the ans\"atze to the first four gates to be the optimal balance of energy resolution from ansatz expressivity and noise. These are the circuits indicated by the final row for each superpotential in \tab{tab:AVQE-Summary}. Truncating the ans\"atze also enables us to effectively extrapolate to larger $\La$.

\section{IBM Hardware Results}\label{HR}
In the previous section we used the AVQE to construct optimal circuits for the various superpotentials and $\La$. We now explore the VQE algorithm on real IBM devices using these problem-tailored ans\"atze. Due to limited available QPU time we are restricted on both the number and size of the systems that we can run on a real device. Despite this, the aim of the initial study is to estimate how the resource cost scales for each superpotential and as we increase $\La$. \tab{tab:vqe-device} is a summary of preliminary results. We study each superpotential with \La ranging from 2 to 8. The ans\"atze used are the ones shown in \tab{tab:AVQE-Summary} and the number of variational parameters is indicated by $N_{\theta}$. 
$|\Delta E|$ is the difference between the VQE energy from the real device and the energy from exact diagonalization. Unlike our classical simulations we no longer see an accuracy better than roughly $10^{-3}$, even for the HO superpotential where classical simulation produces agreement up to machine precision. This is expected and is due to gate noise and readout error within the real device.   
The scaling of this device error can be seen as we increase $\La$. By increasing \La we also increase the number of Paulis $N_P$ and therefore gates required to decompose the Hamiltonian, which in combination with more parametrized gates $N_{\theta}$ within the ansatz causes the QPU usage to increase. We can also see that an increase in $N_{\theta}$ has more impact on usage than an increase in Paulis. For example, in the HO case where $N_{\theta}$ is fixed there is very little change in the QPU usage as $N_P$ increases. In the AHO case where $N_{\theta}$ increases from 1 to 3 we see a significant increase in usage. Additionally, we also observe a clear decline in accuracy.

\begin{table}
\centering
\small
\setlength{\tabcolsep}{4pt}
\renewcommand{\arraystretch}{1.15}
\resizebox{0.8\textwidth}{!}{%
\begin{tabular}{lrrrrrrrrrrr}
\toprule
$W(\qhat)$ & $\La$ & $N_{\theta}$ & $N_P$ & RL & Shots & Iters & Usage/job (s) & Total Usage (s) & $E_{\text{VQE}}$ & $E_{\text{exact}}$ & $|\Delta E|$ \\
\midrule
HO  & 2 & 1 &  2 & 0 &  4096 &  27 &  3 &   81 &  0.0044 & 0.0000 & 0.0044 \\
HO  & 4 & 1 &  4 & 0 &  4096 &  28 &  3 &   84 &  0.0520 & 0.0000 & 0.0520 \\
HO  & 8 & 1 &  8 & 0 &  4096 &  30 &  3 &   90 &  0.0654 & 0.0000 & 0.0654 \\
HO  & 8 & 1 &  8 & 2 &  4096 &  31 &  14 &  434 & -0.0141 & 0.0000 & 0.0141 \\
\midrule
AHO & 2 & 1 &  2 & 0 &  4096 &  30 &  3  &   90 & -0.4296 & -0.4375 & 0.0079 \\
AHO & 4 & 1 & 10 & 0 &  4096 &  30 &  4  &  120 & -0.2665 & -0.1648 & 0.1017 \\
AHO & 8 & 3 & 34 & 0 &  4096 &  76 &  7  &  532 &  6.1470 & 0.0320 & 6.1150 \\
AHO & 8 & 3 & 34 & 0 & 10000 &  94 &  14 & 1316 &  4.3548 & 0.0320 & 4.3228 \\
AHO & 8 & 3 & 34 & 1 &  4096 &  96 &  17 & 1632 &  0.8930 & 0.0320 & 0.8610 \\
AHO & 8 & 3 & 34 & 2 &  4096 &  86 &  28 & 2408 & -0.0359 & 0.0320 & 0.0679 \\
\midrule
DW  & 2 & 1 &  4 & 0 &  4096 &  28 &  7 &  196 &  0.4028 & 0.3572 & 0.0456 \\
DW  & 4 & 3 & 14 & 0 &  4096 & 107 &  7 &  749 &  0.9479 & 0.9066 & 0.0413 \\
DW  & 4 & 3 & 14 & 2 &  4096 &  69 & 28 & 1932 &  0.9231 & 0.9066 & 0.0165 \\
\bottomrule
\end{tabular}
}
\caption{Preliminary VQE and QPU usage results using the real ibm\_torino and ibm\_kingston quantum devices.}
\label{tab:vqe-device}
\end{table}

Not surprisingly, for meaningful results we must implement error mitigation techniques.
For the AHO superpotential with $\La=8$ we tested increasing the number of shots from 4096 to 10000.  This doubled the QPU usage per job with minimal effect on accuracy, suggesting QPU time is better spent on error mitigation rather than increasing shots.
We explore error mitigation using Qiskit's built-in resilience levels (indicated by RL in \tab{tab:vqe-device}) when executing the quantum circuits. Resilience level 0 uses no error mitigation while level 1 provides Twirled Readout Error eXtinction (TREX) measurement twirling and level 2 additionally implements zero-noise extrapolation and gate twirling. For all superpotentials we see an improvement in VQE accuracy when increasing the resilience level, especially in the AHO $\Lambda=8$ case where we see roughly $7\times$ improvement for RL=1 and over $90\times$ improvement for RL=2. However, utilizing error mitigation comes at a significant cost in QPU usage per job, which increases by about $2.5\times$ for RL=1 and about $4\times$ for RL=2. Unfortunately, this makes running even small systems with error mitigation expensive.  Significant increases in QPU resources would be needed to analyse systems with larger $\La$, $N_P$, $N_{\theta}$ and optimizer iterations.
We also observed a significant overhead involved with running the quantum circuits. Due to this overhead, even for the HO case where there is a single gate and only two Paulis we need roughly three seconds QPU usage per job.

\section{Conclusions \& Outlook}\label{Conclusion}
We have reported on our ongoing investigations of spontaneous supersymmetry breaking in supersymmetric quantum mechanics (SQM) using variational quantum algorithms, building on Ref.~\cite{kerfoot2025} by running on real quantum devices. SQM is an ideal test bed since it exhibits a severe sign problem in Monte Carlo formulations, and offers a clear diagnostic for supersymmetry breaking through measurement of the ground-state energy.

By utilizing our AVQE algorithm with a small expressive operator pool we are able to substantially reduce circuit complexity compared to hardware-efficient ans\"atze. Additionally, we observed systematic patterns in the circuit constructions produced as \La increases. Since the initial gates have the largest impact on the energy, we proposed truncated problem-tailored ans\"atze as a practical compromise between expressivity, computational cost and noise. Fewer variational parameters and reduced depth can outweigh the loss in expressivity when noise is present making such circuits more viable for current NISQ hardware. Furthermore, by implementing this truncation we are able to clearly extrapolate ans\"atze to larger $\La$.

Our preliminary runs on IBM quantum hardware highlighted key limitations. Even for small systems ($\La \le 8$) we observe large device-induced energy errors and cannot reach the same accuracy $\lesssim 10^{-3}$ achieved in classical simulations. Built-in error mitigation (resilience levels) can significantly improve energies in some cases, but at a substantial QPU-usage cost. Together with rapid growth in Hamiltonian Pauli decomposition as \La increases, this makes it challenging to scale our VQE studies to larger truncations within limited QPU budgets.

These results have motivated new directions for our ongoing work.
While we have long been interested in the 1+1D Wess--Zumino (WZ) model, this is far more complex than the SQM case, involving a much larger number of qubits and Pauli strings required to represent the Hamiltonian. We have carried out preliminary tests of the AVQE algorithm for the WZ model and although in smaller systems the algorithm does find the groundstate, as we increase both the number of sites and the cutoff \La the algorithm begins to struggle. With our current access to NISQ hardware VQE analyses of the WZ model are not feasible.  Therefore we are exploring more noise-resilient approaches that require fewer circuit evaluations. One particular method of interest is the Sample-based Krylov Quantum Diagonalization (SKQD) algorithm~\cite{skqd}. In this approach both pre- and post-processing is done on the classical device with only basis state sampling done on the QPU. This helps eliminate the poor energy resolution we have seen with the VQE as a result of gate and readout errors. SKQD requires that the initial state has a non-zero overlap with the groundstate and therefore we may see a potential improvement in SKQD convergence by first using the AVQE to find good ans\"atze. Even if the ans\"atze returned by the AVQE are not optimal (as we have seen in this work) we can truncate and be confident that we have a significant overlap with the groundstate before performing the SKQD analysis.  Our preliminary SKQD tests for the WZ model are producing promising results which we will present in a future publication.

\vspace{8 pt}
\noindent \textbf{\textsc{Acknowledgments:}} JK was supported through the Liv.Inno Centre for Doctoral Training, funded by the UK's Science and Technology Facilities Council (STFC) under grant agreement {ST/W006766/1}.
EM and DS were supported by UK Research and Innovation (UKRI) Future Leader Fellowship {MR/X015157/1}, with additional support for DS from STFC consolidated grant {ST/X000699/1}.
Classical numerical calculations were carried out at the University of Liverpool.
We acknowledge the use of IBM Quantum services for this work, provided via the UK National Quantum Computer Centre [NQCC200921], which is a UKRI Centre and part of the UK National Quantum Technologies Programme.

\vspace{8 pt}
\noindent \textbf{Data Availability Statement:} All classical simulation data used in this work is available through the open data release Ref.~\cite{data}. Preliminary quantum device data can be provided upon reasonable request.

\begingroup
\setlength{\bibsep}{0pt}
\setlength{\itemsep}{0pt}
\setlength{\parskip}{0pt}
\setlength{\parsep}{0pt}
\bibliographystyle{utphys}
\bibliography{Bib}
\endgroup

\end{document}